\documentclass[10pt]{article}
\usepackage[utf8]{inputenc}
\usepackage{multicol}
\usepackage[margin=0.67in]{geometry}
\usepackage[T1]{fontenc}
\usepackage{mathptmx}
\usepackage{titlesec}
\usepackage{indentfirst}
\usepackage{graphicx}
\usepackage[affil-it]{authblk} 
\usepackage{etoolbox}
\usepackage{lmodern}
\usepackage{xurl}
\usepackage{authblk}
\makeatletter
\usepackage{enumitem}
\usepackage[table]{xcolor}
\patchcmd{\thebibliography}{\section*{\refname}}{}{}{}
\patchcmd{\@maketitle}{\LARGE \@title}{\fontsize{24}{27}\selectfont\@title}{}{}
\usepackage{amsmath} 
\usepackage[utf8]{inputenc}
\makeatother
\usepackage{titling}
\usepackage{enumitem}
\setlist[itemize]{noitemsep}

\predate{}
\postdate{}

\title{NFT Wash Trading Detection \vspace{-0.5em}}
\author{\large Derek Liu, Francesco Piccoli, Katie Chen, Adrina Tang, and Victor Fang Ph.D. \thanks{Liu, Piccoli, Fang are affiliated with AnChain.AI Inc, San Jose, California 95113. (Derek.Liu@Anchain.ai); 
Chen is affiliated with Columbia University, New York, NY 10032; 
Tang is affiliated with Massachusetts Institute of Technology, Cambridge, MA 02139;
Chen and Tang work on this research project during their data science summer internship at AnChain.AI. 
} \\ \textup{AnChain.AI, San Jose, California} \\ \textup{MIT} \\ \textup{Columbia University} \vspace{-1.0em}}
\date{}
\raggedcolumns

\begin{document}
\maketitle

\titleformat*{\section}{\normalfont\scshape\filcenter}
\begin{multicols}{2}
\small\textbf{\textit{Abstract}} --  Wash trading is a form of market manipulation where the same entity sells an asset to themselves to drive up market prices, launder money under the cover of a legitimate transaction, or claim a tax loss without losing ownership of an asset. Although the practice is illegal with traditional assets, lack of supervision in the non-fungible token (NFT) market enables criminals to wash trade and scam unsuspecting buyers while operating under regulators' radar. AnChain.AI designed an algorithm that flags transactions within an NFT collection’s history as wash trades when a wallet repurchases a token within 30 days of previously selling it. The algorithm also identifies intermediate transactions within a wash trade cycle. Testing on 7 popular NFT collections reveals that on average, 0.14\% of  transactions, 0.11\% of wallets, and 0.16\% of tokens in each collection are involved in wash trading. These wash trades generate an overall total price manipulation, sales, and repurchase profit (defined below) of \$900K, \$1.1M, and -\$1.6M respectively. The results draw attention to the prevalent market manipulation taking place and inform unsuspecting buyers which tokens and sellers may be involved in criminal activity.   
\medskip

\textbf{\textit{Index Terms}} -- NFT, cryptocurrency, anti-money laundering, tax evasion, market manipulation, wash trading, blockchain, smart contracts, web3, cybersecurity
\section*{Nomenclature}
\normalsize\noindent \textbf{NFT} - Non-fungible token\\
\textbf{ETH} - Ethereum\\
\textbf{BFS} - Breadth-first search \\
\textbf{MANA} - Ethereum token used to purchase virtual land in Decentraland\\
\textbf{Collection} - A group of related NFT tokens, typically from the same creator\\
\textbf{PM} - Price manipulation\\
\textbf{Opensea} - A popular NFT marketplace
\section*{I. Introduction} 
\par Non-fungible tokens (NFTs) are unique cryptographic assets with records on a decentralized blockchain, making it difficult for their ownership to be faked. NFTs were first introduced to the public in 2014, and surged in popularity in 2021 as digital assets grew in mainstream attention [1], [2]. NFTs can take the form of an endless variety of objects: digital art, video game collectibles, music, etc [1], [2]. 
\par The booming NFT marketplace also comes with exposure to financial risks, such as wash trading [2]. Wash trading occurs when the same person or entity sells and buys back the token within a short period of time and was rampant amongst stock market manipulators before the 1930s, when traders would collude to artificially drive up stock prices and short those stocks for large profits [3], [4].
\par  In 1936, deliberate wash trading to manipulate the marketwas outlawed with the Commodity Exchange Act and the Securities Exchange Act. A Wash Sale Rule was also established, officially classifying the buy back of a “substantially identical” stock within 30 days of selling it as a wash sale and banning any tax loss claims resulting from wash sales  [5]. Although conducting a wash sale without claiming tax losses is not strictly illegal, egregious behaviors will lead to further inspection and borderline the illegal practice of wash trading. 
\par A key difference between NFTs and stocks to note: certain tokens within an NFT collection may vary in value significantly, so this analysis treats each specific token as “substantially identical” only to themselves.
\par Although rules against wash trading designed specifically for NFTs do not currently exist, existing commodity and security laws may apply [3]-[4]. The Commodity Exchange Act in 1936 prohibits making transactions that appear as sales but do not change the trader’s market position or induce market risk [4], [6]. The Securities Exchange Act in 1934 empowered the Securities Exchange Commission (SEC) to regulate all aspects of the securities industry and advance the transparency and accuracy of financial information in the market [7]. These rules currently may not be enforced to the fullest extent and are subject to loopholes in the NFT market, as NFTs are a relatively new phenomenon and there is a still active debate over which asset class they belong to and which regulations apply to them or not. 

\par Due to the largely unregulated NFT marketplace, the prevalence of wash trading poses a substantial threat to the integrity of the market and to unsuspecting buyers who are unaware that certain token prices and trading volumes have been artificially manipulated [2]-[4]. 
\par Illicit wash trading may be attractive for multiple reasons, the most common of which are price inflation, money laundering, and tax loss harvesting [3]. 
\par Trading the same token repeatedly between multiple wallets can artificially inflate the token’s transaction volume and thus create the appearance of high demand and elevated market value, allowing the owners of these tokens to make a large profit when they sell the token at the inflated price [3], [4]. 

\par In addition, laundering money using NFTs is attractive to criminals as NFTs can be sold pseudo-anonymously online without needing to physically move assets. Financial criminals might conduct wash trades to move large sums of money under the cover of legitimate NFT transactions, or to indirectly make payments to third-party brokers. The volatile price changes in the crypto market also help criminals avoid suspicion with large money transfers. Regulators and law enforcement are also currently insufficiently equipped with effective tools at identifying and monitoring criminals exploiting this new technology.

\par   NFTs can also be sold and repurchased at a loss to exploit loopholes in taxation law, enabling traders to claim a potential tax loss without actually losing ownership of the token [5]. This strategy is especially prevalent towards the end of the tax year or during market dips [5], [8]. 
\par With the incentives to wash trade illustrated above, the purpose of this analysis is to create an algorithm that detects wash trading activity and to inform unsuspecting buyers which tokens may have artificially inflated prices. The project aims to build on top of previous work in the field by further examining the profits made by suspected wash traders, thereby placing more scrutiny on price manipulators, money launderers and tax evaders [2]. 

\section*{II. Methodology}
\noindent \textit{A. Data Collection}
\par The data used in this project is procured from Ethereum, a decentralized blockchain with transaction records that are available to the public. Senders and receivers in these transactions are user-created Ethereum accounts, and only sales from OpenSea were considered in this project. The following information regarding transfer and sale transactions in the history of Art Blocks, Azuki, Bored Ape Yacht Club, Decentraland, Doodles, Mutant Ape Yacht Club, and Otherdeed NFT collections from launching to June 2022 were assembled and utilized:
\begin{itemize}
\item Transaction sender address
\item Transaction receiver address
\item Transaction timestamp
\item Transaction hash
\item NFT collection name
\item NFT collection token identifier
\item Cryptocurrency amount transacted (if applicable)
\item Cryptocurrency token used for payment
\item Approximate USD value conversion for cryptocurrency payment
\end{itemize}
\medskip\noindent \textit{B. Detecting and Graphing Wash Sales}
\par In order to detect instances of wash trading for each token in the data, it was necessary to group transactions by token and parse through each token’s sale history. 
\par The algorithm loops through each token’s sale history to find transactions where a sender wallet in one transaction is subsequently the recipient of the same token in another transaction. If the repurchase occurs within 30 days of the original sale, it is flagged as breaking the “Wash Sale Rule” that currently applies to commodity and security trading [3]. 
\par An additional layer of complexity is introduced in NFT smart contract design which allows wallets to transfer NFT tokens without exchanging for monetary value. To tackle this problem the algorithm creates two directed graphs for each token:
\begin{itemize}
\item A token transfer graph visualizes all sales and transfers of a particular token including those where no money is exchanged (Fig. 1) 
\item Token sale graph visualizes all sales of a particular token where the token was exchanged for money (Fig. 2)
\end{itemize}
\par Each numbered node in both graphs corresponds to a wallet address appearing in the token’s transaction history, and each edge corresponds to a transaction moving the token from one wallet owner to another. Node sizes are proportional to the number of edges the node is connected to, meaning the number of transactions the wallet takes part in.

\begin{table*}[b]
\begin{center} \footnotesize TABLE I\\
\smallskip
 Sale transaction history of BAYC ID 1332. Wallet address highlighted in red is the wash trading wallet\\
 \smallskip
  \begin{tabular}{|c|c|c|c|}
\hline
\rowcolor{gray!25}
Timestamp & \footnotesize Source & Dest & Value (\$) \\
\hline
2021-06-01 0:49:43	& 0xdc82142e5fa1ad18bee3f351d0c820db63ca5a91 & \cellcolor{red!15} 0x1729ae0e8f58d55de0f209273759cb644405478a & 5124.66\\
\hline
2021-06-20 1:41:46	& \cellcolor{red!15} 0x1729ae0e8f58d55de0f209273759cb644405478a	& 0x30f0149363f860bd37015a77da1db8b5845545cc	& 8503.60\\
\hline
2021-07-10 17:53:09	& 0x30f0149363f860bd37015a77da1db8b5845545cc &	0xc91b761085e6d9059e1e5012cc82eec9ec3110fc &	9239.76\\
\hline
2021-07-17 19:41:02 &	0xc91b761085e6d9059e1e5012cc82eec9ec3110fc	& \cellcolor{red!15} 0x1729ae0e8f58d55de0f209273759cb644405478a  &	16932.51\\
\hline
2021-08-21 10:52:50	& \cellcolor{red!15} 0x1729ae0e8f58d55de0f209273759cb644405478a	& 0x8f18d6a49bb392a84a4a4c03b69d29179e333946 &	75425.67\\
\hline
\end{tabular}
\end{center}
\end{table*}
\normalsize 

\begin{center}
\includegraphics[scale=.5]{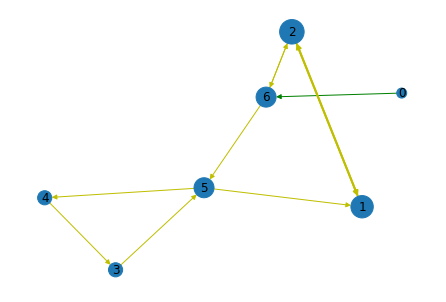}
\end{center}
\footnotesize \noindent Fig. 1.  Graph representing all sale and transfer transactions for Azuki Token ID 9845. Yellow edges represent transactions in a cycle. Green edges represent other regular transfers.

\begin{center}
\includegraphics[scale=.5]{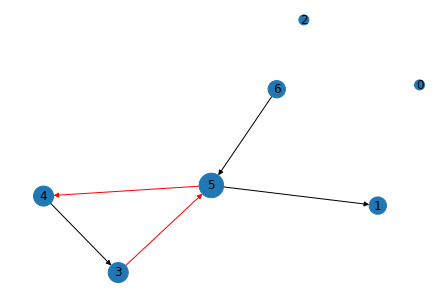}
\end{center}
\footnotesize \noindent Fig. 2.  Graph representing only sale transactions for Azuki Token ID 9845. Red edges represent flagged wash sales. Black edges represent all other sales.

\normalsize \medskip

\par When analyzing a graph it is important to clarify the following terms:
\begin{itemize}
\item Cycle - a group of transactions starting from when a seller first sells an NFT token and ending when the same seller buys back the same NFT token. Any transaction in between the original sale and the repurchase will also be included in the cycle as an intermediary.
\item Repurchase -  the transaction buying back the same NFT token. In other words, this will always be the last transaction in a cycle.
\item Wash sale -  transactions where a seller sells an NFT token and buys back the same NFT token within 30 days. Wash sales will always include a repurchase.
\item Intermediary - any edge in a wash sale cycle which is not a wash sale.
\end{itemize}

\begin{center}
\includegraphics[scale = 0.5]{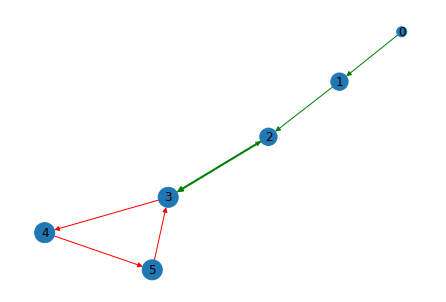}
\end{center}
\footnotesize \noindent Fig. 3.  Graph representing all transfer and sale transactions for Bored Ape Yacht Club ID 6946. Red edges are transactions in a wash sale cycle. Green edges are other transfers.

\normalsize \medskip
\par In Fig. 3, the entire \textbf{cycle} is denoted by the red edges (3 → 4, 4 → 5, 5 → 3). The \textbf{repurchase} occurs on edge (5 → 3). If (5 → 3) occurs within 30 days of (3 → 4), then the repurchase edge and the first sale edge (3 → 4) will also be flagged as a \textbf{wash sale}. The \textbf{intermediary} in the above graph is the (4 → 5) edge.
\medskip

\noindent \textit{C. Identifying Intermediaries}
\par Once the algorithm identifies the wallets and transactions that strictly break the “Wash Sale Rule,” it looks for other intermediary wallets and transactions that are involved in the wash sale cycle.
\par In Fig. 1, wallet 5 sells a token to wallet 4, who then sends the token to wallet 3, who sells it back to wallet 5 within 30 days of the first sale. The sale between wallets 4 and 3 is another suspicious transaction that does not directly break the “Wash Sale Rule.” In the above example, wallet 5 is the prime suspect for being a wash seller, while wallets 4 and 3 are also suspicious wallets of interest for their involvement in the wash sale cycle. 
\par The algorithm’s solution to identify these intermediary wallets and transactions uses a breadth-first search (BFS) algorithm that looks for a path in the graph starting with the first wash trading sale and ending with the buy-back sale. All intermediary wallets and transactions within the path, regardless of transfer or sale, are then flagged for suspicion of involvement.
\medskip

\noindent \textit{D. Tracking Changes in Price}
\par Using the labeled data from the graph algorithm, wash trade cycles can also be plotted on a time series with values for any specific token. A large gain/loss in a token’s value after wash selling resulting in abnormal prices compared to other sales in the collection can provide a valuable signal for risk.
\medskip

\noindent \textit{E. Determining Profit}
\par To calculate profits made from the identified wash sales, the algorithm parses through the sale transaction history of each token with wash sale activity.
\par In traditional finance, profit is commonly determined by subtracting the cost of procuring an asset from the value it generated. In this project, three types of profits are calculated for different purposes (to be defined in further detail below): 
\begin{itemize}
    \item price manipulation profit
    \item sale profit
    \item repurchase profit
\end{itemize}

\par The transaction history for BAYC token ID 1332 can be taken as an example in Table I, with a wash sale cycle taking place in rows 2-4, when wallet \textbf{0x1729ae0e8f58d55de0f209273759cb644405478a} sells the token for \$9K (row 2, Table I) and buys it back less than 30 days later for \$17K (row 4, Table I).

\par The final results for price manipulation, sales, and repurchase profit are \$70K, \$3K, and \$58K respectively as shown in Table II. 

\begin{center} \footnotesize TABLE II\\
\smallskip
Profit data for BAYC ID 1332\\
\smallskip
\footnotesize
\begin{tabular}{|c|c|c|}
\hline
\rowcolor{gray!25}
\footnotesize PM Profit (\$) & Sales Profit (\$) & Repurchase Profit (\$)\\
\hline
70,301.00 & 3,378.94 & 58,493.16 \\
\hline
\end{tabular}
\end{center}
\normalsize

\par \textit{1) Price Manipulation Profit:}
\par  Price manipulation profit is calculated as: \\ \\
price of token exiting cycle - price of token entering cycle \\

\par BAYC Token 1332’s transaction history containing the wash sale cycle is shown in Table 1, and rows 1 and 5 are used in the price manipulation calculation. In this example, the price of the token entering the cycle (\$5K) is subtracted from the price of the token exiting the cycle (\$75K), resulting in a total of \$70K price manipulation profit. 
\par A key assumption when using this profit calculation is that the wallets within the cycle are colluding with each other to inflate the price. This assumption is made because a manual review of outliers showcases that it is likely for the addresses within cycles of short time periods to be owned by the same individual or group of individuals. Following this assumption, the price manipulation profit method does not take into account any of the purchase prices within the cycle itself. 
\par The purpose of this profit calculation is to identify the wash sale cycles which have demonstrated extremely high rises in price within a short timeframe. This method may be an effective way to capture sellers who work together to artificially increase the price of an NFT token by selling amongst each other and deceiving buyers with a false sense of demand. 

\medskip
\par \textit{2) Sale Profit:}
\par  Sale profit is calculated as: \\ \\
token sale price - previous token purchase price \\
\par Following the BAYC Token 1332 example shown in Table I, the sales profit is calculated from rows 1 and 2 by subtracting the price that the wash trading wallet bought the token for before the cycle (\$5.1K) from the amount the wallet first sold the token for (\$8.5K), resulting in a total of \$3.4K profit from the sale.
\par The last token purchase price is not necessarily the value the seller bought the token for, if the seller received the token as a transfer free of cost. This profit calculation assumes that the recipient of free transfers is related to the previous buyer, and the cost from the previous buyer is carried over to the eventual sale.
\par The purpose of this profit calculation is to identify outliers with extremely high gain or loss in a single transaction. Sale profit outliers with large losses may indicate heightened suspicions on tax evasion while large gains may indicate attempts to manipulate price.

\medskip
\par \textit{3) Repurchase Profit:}
\par  Repurchase profit is calculated as: \\ \\
post-cycle token sale price - token repurchase price \\
\par Using the example in Table I, this profit is calculated using rows 4 and 5 by subtracting the token repurchase price  (\$17K) from the amount of the post-cycle token sale price (\$75K), resulting in a \$58K profit from the sale.
\par Repurchase profit relies on a similar assumption as sale profits. The next token sale is not necessarily made by the repurchase wallet if the repurchase wallet transfers the token to another wallet free of cost. Assuming that donors of free transfers are related to the recipient who is the next seller, then the cost from the next sale is used in the calculation. 
\par The purpose of this profit calculation is similar to that of the sale profit calculation, just capturing the signal at the end of the wash sale cycle. Large losses may indicate heightened suspicions on tax evasion while large gains may indicate attempts to manipulate price.

\section* {III. Results}

\par The detection algorithm was run on seven different NFT collections in this analysis:
\begin{itemize}
    \item Art Blocks
    \item Azuki
    \item Bored Ape Yacht Club (BAYC)
    \item Decentraland
    \item Doodles
    \item Mutant Ape Yacht Club (MAYC)
    \item Otherdeed
\end{itemize}
\medskip
\begin{center} \footnotesize TABLE III\\
\smallskip 
Wash Sale Statistics for 7 NFT collections\\
\smallskip
\begin{tabular}{|>{\columncolor{gray!25}}c|c|c|}
\hline
\rowcolor{gray!25}
 Collection & \# Wash Sales & \% Wash Sales \\
\hline 
ArtBlocks &  140 & 0.09 \\ 
\hline
Azuki & 26 & 0.1  \\
\hline
BAYC & 72 & 0.275 \\
\hline
Doodles & 22 & 0.098 \\
\hline
Decentraland & 6 & 0.141 \\
\hline
MutantApe & 52 & 0.161 \\
\hline
Otherdeed & 29 & 0.085 \\
\hline
Average & 49.571 & 0.136 \\ 
\hline
Total & 347 & 0.115\\
\hline
\end{tabular}
\end{center}
\bigskip\bigskip\bigskip\bigskip

\begin{center} \footnotesize TABLE IV\\
\smallskip 
Wash Token Statistics for 7 NFT collections \\
\smallskip
\begin{tabular}{|>{\columncolor{gray!25}}c|c|c|}
\hline
\rowcolor{gray!25}
 Collection & \# Wash Tokens & \% Wash Tokens \\
 \hline
\footnotesize ArtBlocks & 68 & 0.07 \\
\hline
Azuki & 13 & 0.148 \\
\hline
BAYC & 36 & 0.414 \\
\hline
Doodles & 10 & 0.125 \\
\hline
Decentraland & 3 & 0.106 \\
\hline
MutantApe & 26 & 0.209 \\
\hline
Otherdeed & 7 & 0.027 \\
\hline
Average & 23.286 & 0.157 \\
\hline
Total & 163 & 0.010\\
\hline
\end{tabular}
\end{center} 
\begin{center} \footnotesize TABLE V\\
\smallskip 
Wash Wallet Statistics for 7 NFT collections \\
\smallskip
\begin{tabular}{|>{\columncolor{gray!25}}c|c|c|}
\hline
\rowcolor{gray!25}
 Collection & \# Wash Wallets & \% Wash Wallets \\
 \hline
\footnotesize ArtBlocks & 56 & 0.134 \\
\hline
Azuki & 13 & 0.084 \\
\hline
BAYC & 34 & 0.265 \\
\hline
Doodles & 7 & 0.048 \\
\hline
Decentraland & 3 & 0.095 \\
\hline
MutantApe & 26 & 0.107\\
\hline
Otherdeed & 8 & 0.03\\
\hline 
Average & 21 & 0.109 \\
\hline 
Total & 147 & 0.106 \\
\hline
\end{tabular}
\end{center}

\begin{center} \footnotesize TABLE VI\\
\smallskip 
Price Manipulation Profit Statistics for 7 NFT collections \\
\smallskip
\begin{tabular}{|>{\columncolor{gray!25}}c|c|c|c|}
\hline
\rowcolor{gray!25}
 Collection & Max PM (\$) & Avg PM (\$) & Total PM (\$) \\
 \hline
\footnotesize ArtBlocks &  413,519.49  & 827.23 & 56,251.60\\
\hline
Azuki & 31,946.96 & -3,859.27 & -50,170.51 \\
\hline
BAYC & 344,448.96 & 20,964.53 & 754,722.91\\
\hline
Doodles & 67,635.82 & -1,161.98 & -11,619.81 \\
\hline
Decentraland & 15,992.68 & 10,092.20  & 30,276.59 \\
\hline
MutantApe & 73,715.66 & 4,266.59 & 110,931.25 \\
\hline
Otherdeed & 77,963.65 & 5,728.85 & 40,101.93 \\
\hline
Overall & 413,519.49 & 5,431.45 & 930,493.96 \\ 
\hline
\end{tabular}
\end{center}

\begin{center} \footnotesize TABLE VII\\
\smallskip 
Sales Profit Statistics for 7 NFT collections \\
\smallskip
\begin{tabular}{|>{\columncolor{gray!25}}c|c|c|c|}
\hline
\rowcolor{gray!25}
 Collection & Max SP (\$) & Avg SP (\$) & Total SP (\$) \\ 
 \hline
 Art Blocks & 84,886.29  & -3,469.34 & -235,915.24\\
\hline
\footnotesize Azuki &  39,635.33  & 5,296.77 & 68,858.05\\
\hline
BAYC & 286,114.18 & 18,094.01 & 651,384.41 \\
\hline
Doodles & 34,559.06 & -5,589.61 & -55,896.13\\
\hline
Decentraland & 14,731.92 & 5,428.83  & 16,286.48  \\
\hline
MutantApe & 80,278.14 & 19,622.41 & 510,182.67 \\
\hline
Otherdeed & 143,000.73 & 22,217.56 & 155,522.93 \\
\hline
Overall & 286,114.18 & 9,295.71& 1,110,423.17 \\ 
\hline
\end{tabular}
\end{center}

\begin{center} \footnotesize TABLE VIII\\
\smallskip 
Repurchase Profit Statistics for 7 NFT collections \\
\smallskip
\begin{tabular}{|>{\columncolor{gray!25}}c|c|c|c|}
\hline
\rowcolor{gray!25}
 Collection & Max RP (\$) & Avg RP (\$) & Total RP (\$) \\ 
 \hline
 Art Blocks & 98,293.57  & -1,391.88 & -94,647.83 \\
\hline
\footnotesize Azuki & 9,647.09 & -4,633.43 & -60,234.59 \\
\hline
BAYC & 110,632.83 & -11,542.34 & -415,524.10 \\
\hline
Doodles & 67,657.83 & 4,363.30 & 43,632.97 \\
\hline
Decentraland & 17,524.91 & 1,387.36  & 4,162.09  \\
\hline
MutantApe & 46,044.20 & -9,556.59 & -248,471.44\\
\hline
Otherdeed & 457.07 & -116,468.91 & -815.282.36 \\
\hline
Overall & 110,632.83 & -19,029.87 & -1,586,365.26\\ 
\hline
\end{tabular}
\end{center}
\medskip
\par The amount and percentage of wash sales, tokens, and wallets in each collection are aggregated and profits are totaled in Tables III-VIII. These statistics take into account sale transactions where money was exchanged; transfers are excluded. 
\par Over all collections that the algorithm was tested on, on average, 0.136\% of  transactions, 0.157\% of tokens, and 0.109\% of wallets in each collection are involved in wash trading (row 8, Table III-V). These wash sales generate an overall total price manipulation, sales, and repurchase profit of around \$900K, \$1.1M, and -\$1.6M respectively (row 8, Tables VI-VIII).
\par Comparing the statistics between different collections reveals that the Bored Ape Yacht Club (BAYC) dataset has the highest percentage of wash traded transactions, tokens, and wallets (Tables III-V). 0.275\% of sale transactions, 0.414\% of tokens, and 0.265\% of wallets in the collection’s transaction history take part in wash trading (row 3, Tables III-V). Furthermore, wash trading in the BAYC collection accumulates total price manipulation, sales, and repurchase profits of \$755K, \$651K, and -\$416K respectively (row 3, Tables VI-VIII). 
\par This prompts a deeper analysis into the trading of BAYC tokens. Fig. 7 highlights which periods of time in the past 2 years had a high amount of wash traded transactions in the BAYC collection, and many of the peaks happened in mid to late 2021.
\par Other collections to take a closer look at from this analysis are Art Blocks, which contained 68 tokens with detected wash trading activity and a max price manipulation profit of \$414K, as well as Mutant Ape which had the second highest percentage of wash traded transactions at 0.161\% (Table III, IV, VI). 

\begin{center}
\includegraphics[scale = 0.5]{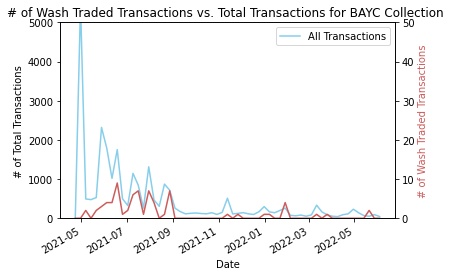}
\end{center}
\footnotesize\noindent{Fig. 7.  Time series graph for number of wash traded vs. total transactions in Bored Ape Yacht Club collection}

\normalsize
\section* {IV. Discussion}
\par This analysis provides a method for utilizing public information gathered from the Ethereum blockchain to detect wash trades, calculate financial gains, and benchmark statistics on a few of the most popular NFT collections to date. It is insightful to explore a few case studies on different outliers captured by the three methods of profit calculation at the token level.
\medskip
\par \noindent \textit{A. Abnormal token value changes}
\par When an abnormally large gain in value occurs for a token after a wash cycle, it may indicate that the price of the token has been manipulated. In Fig. 8, the price histories of BAYC tokens 8099 and 8498 are plotted on a time series and compared against the average value of sales in the same collection during the same month. 
\par Token 8099 experienced a wash cycle with a maximum value of \$166K in August 2021 when the average collection sale price was \$66K. Breaking the cycle in November 2021, the same token is sold at a 67\% gain for \$276K compared to an average collection sale price of \$212K.
\par Token 8498 had its first wash sale transaction with a fair market price value of \$12K in July 2021. Breaking the cycle in August 2021, the same token is sold at a 1400\% gain for \$180K.
\par Both examples illustrate rapid increases in price that are outside of typical collection sale prices, indicating high likelihood that the price of the token has already been successfully manipulated.

\begin{center}
\includegraphics[scale = 0.22]{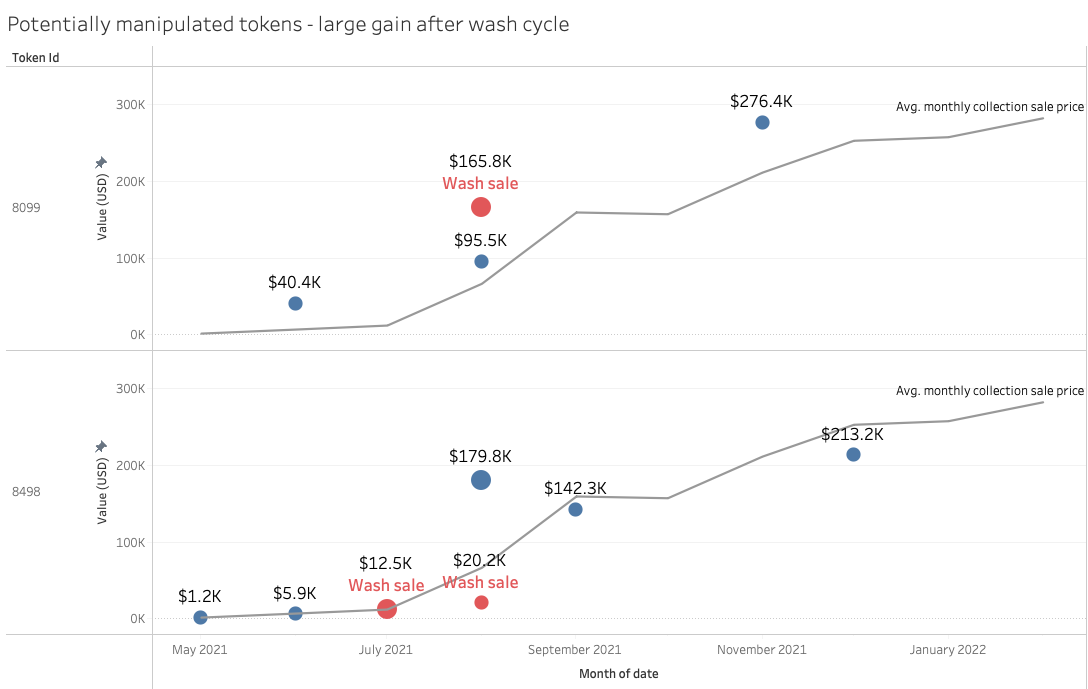}
\end{center}
\footnotesize\noindent Fig. 8.  Tracking sale value of BAYC Token 8099 and 8498 over time. Dots in red indicate a wash sale. Line indicates the average collection sale price in the same month.

\medskip
\normalsize \par Similarly, rapid increase in token’s value brought forth by wash cycles with no exits may hold predictive importance for indicating when price manipulations are being attempted. In Fig. 9, the price histories of BAYC tokens 5862 and 8259 are plotted on a time series and compared against the average value of sales in the same collection during the same month.
\par Token 5862 was first purchased at the average collection sales price in August 2021, but entered a wash cycle with a maximum value of \$194K in August 2021 which far exceeded the average collection price of \$66K.
\par Token 8259 was also purchased near the average collection sales price of \$206K in November 2021, and experienced a wash cycle with a maximum value of \$271K in November 2021 representing a 32\% price increase.
\par Both examples illustrate rapid increases in price during the wash sale cycle that are outside of typical collection sale prices. Since these tokens have not yet been sold they are at high risk of deceiving buyers into purchasing well above market value.

\begin{center}
\includegraphics[scale = 0.22]{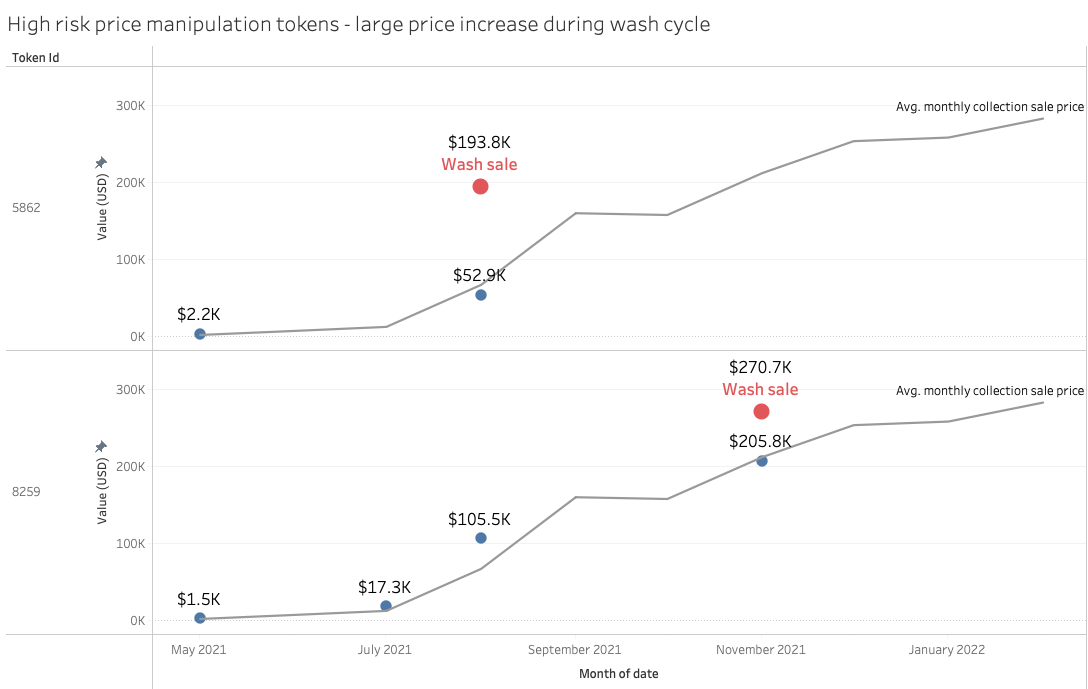}
\end{center}
\footnotesize \par \noindent Fig. 9.  Tracking sale value of BAYC tokens 5862 and 8259 over time. Dots in red indicate a wash sale. Line indicates the average collection sale price in the same month.

\medskip
\normalsize \par Large losses resulting from wash cycles serve as strong indicators for wallets looking to seek large tax write offs through tax loss harvesting strategies. In Fig. 10, the price histories of BAYC tokens 1904 and 7856 are plotted on a time series and compared against the average value of sales in the same collection during the same month.
\par Token 1904 was first purchased at the average collection sales price in June 2021 for \$7K, but was sold at no cost in August 2021.
\par Token 7856 was also purchased at the average collection sales price of \$7K in June 2021, and also sold at no cost in August 2021.
\par Although not shown in Fig. 10, it is worth mentioning that both wash sale examples above occur between the same two wallets. Presumably these wallets are related, though it is still possible the owner attempts to report the sale as a loss on tax returns.

\begin{center}
\includegraphics[scale = 0.22]{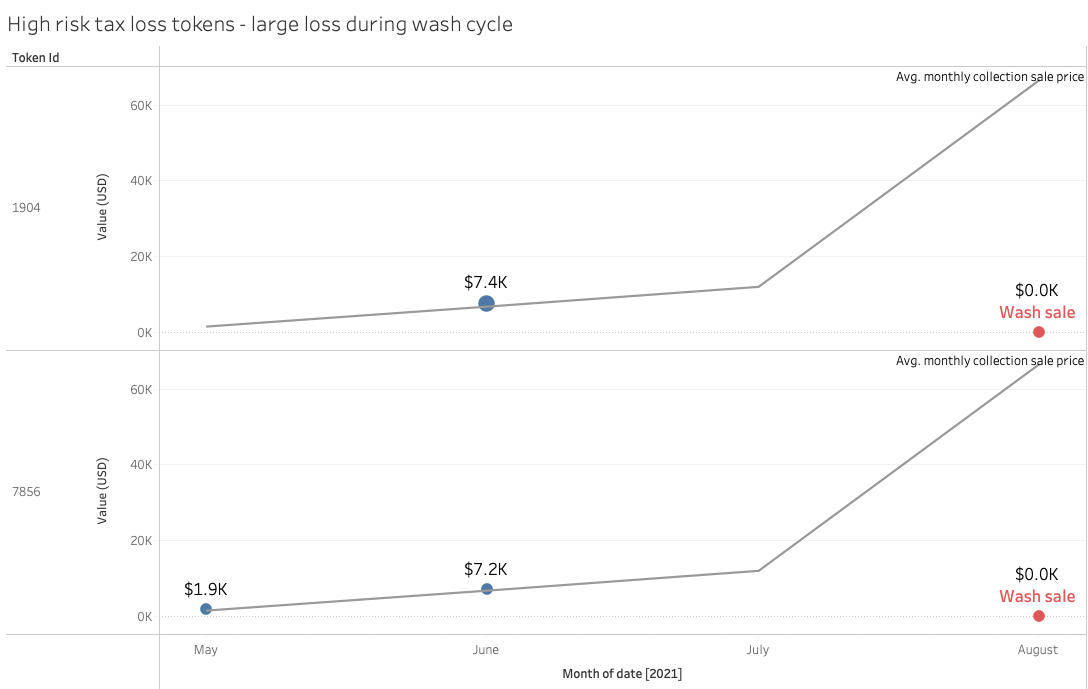}
\end{center}
\footnotesize \noindent Fig. 10.  Tracking sale value of BAYC Token 1904 and 7856 over time. Dots in red indicate a wash sale. Line indicates the average collection sale price in the same month.

\medskip
\normalsize \par \noindent \textit{B. Profit Outlier Analysis}
\par Closely examining the outliers from each method of profit calculation highlights the significant effects of wash trading on certain NFT tokens. Specific token level transaction history where the detection algorithm indicates irregularities in gain/loss are reviewed below to provide insight and further context for assessing risk.
\medskip
\par \textit{1) Price Manipulation Profit:}
\par The top price manipulation gain from wash trading recorded in this analysis was observed in Art Blocks token 78000189, reaching \$423K of value increase in less than one month (Fig. 11).
\par In this example, the wash sale cycle initiated when wallet 8 (address \\ \textbf{0xe1d29d0a39962a9a8d2a297ebe82e166f8b8ec18}) \\ purchases the token on July 31, 2021 for 23.189 ETH (approximately \$59K). The token was sold on August 15, 2021 for 40 ETH (approximately \$132K) and repurchased on August 23, 2021 for 124 ETH (approximately \$412K) concluding the wash sale cycle. On the same day of repurchase, the seller then sold the token to another buyer for 145 ETH (approximately \$482K).
\par Within the month-long timespan containing the wash sale cycle, the price of the NFT token rocketed to over 720\% of the original purchase price.

\begin{center}
\includegraphics[scale = 0.5]{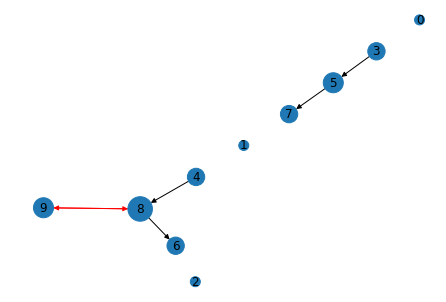}
\end{center}
\footnotesize \noindent Fig. 11.  Sales transactions for Art Blocks ID 78000189. Red edges are transactions in a wash sale cycle.

\medskip
\normalsize \par Another price manipulation profit outlier with lower overall value but higher rate of trading was detected on token ID 55343 in the Otherdeed collection. In this token’s transaction history, 17 wash sales appeared within hours of each other, with 2 wallets trading the token between each other to drive the price from \$14 to \$197 (Fig. 12). The high number of wash sales back and forth is indicative that wallets 1 (address \\ \textbf{0x837e6fd5d2b39b6fd2791ba8a4a364d104c18e15)} \\ and 2 (address \\ \textbf{0x2156001ecebe8fdcd46c0c9be0738d48b2e98d58}) \\ are working together to artificially inflate the price. 

\begin{center}
\includegraphics[scale = 0.5]{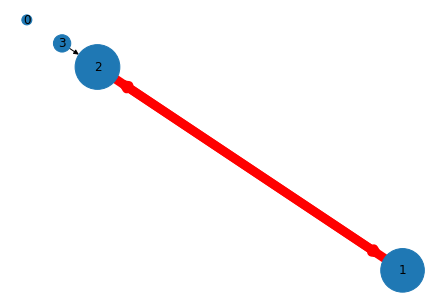}
\end{center}
\footnotesize \noindent Fig. 12.  Sales transactions for Otherdeed ID 55343;  Red edges are transactions in a wash sale cycle. Edges widths correspond with the number of times that the edge is traced.
\medskip
\begin{table*}[b]

\begin{center} \footnotesize TABLE IX:\\
\smallskip 
Top 5 Wallets in Bored Ape Yacht Club ranked from greatest to least sales profit made
\smallskip
\begin{tabular}{|c|c|c|c|c|c|}
\hline \rowcolor{gray!25}
Wallet & Profits(\$) & Wash Sale & PM Profit & Sale Profit & RP Profit\\ 
\hline
0xd3fc6fec4b219c2d74b366fee6b585df71611533 & 8,842,929.84  & TRUE & 5,690.09 & 3,188.85 & 5,813.47 \\
\hline
0xd387a6e4e84a6c86bd90c158c6028a58cc8ac459 & 5,653,667.77 & FALSE & -- & -- & -- \\
\hline
0x31a47094c6325d357c7331c621d6768ba041916e & 4,425,383.87 & FALSE & -- & -- & -- \\
\hline
0xed47015bb8080b9399f9d0ddfc427b9cee2caab1 & 4,302,131.56 & FALSE & -- & -- & -- \\
\hline
0x721931508df2764fd4f70c53da646cb8aed16ace & 4,289,367.18 & FALSE & -- & -- & -- \\
\hline
\end{tabular}
\end{center} 
\end{table*}

\normalsize\par \textit{2) Sale Profit:}
\par A top sale loss with wash sale identified by the algorithm is seen on Azuki token 5105 wallet 10 (address \textbf{0xea4Feb8E55a17EeD317b2804e1F49040d1b43299}) (Fig. 14). 
\par On May 6, 2022 wallet 4 (address \\ (\textbf{0x2cf84928261f655a47d04ec714d3bedf9375de46}) \\ purchased the token for 23.1 WETH (approximately \$62K) and transferred the token to wallet 10 on the following day (Fig. 13). On May 31, 2022 wallet 10 sold the token for just 0.05 WETH (approximately \$97) , realizing a loss of -\$62K before repurchasing a few minutes later for only 0.06 ETH (approximately \$119) (Fig. 14). 
\par The dramatic price deflation in a short period of time indicates suspicious behavior and may lead to exploiting tax evasion loopholes such as tax loss harvesting.

\begin{center}
\includegraphics[scale = 0.55]{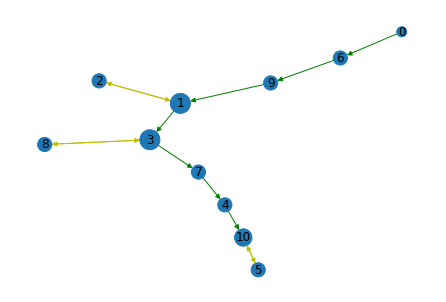}
\end{center}
\footnotesize \noindent Fig. 13.  Transfer transactions for Azuki ID 5105. Yellow edges represent transactions in a cycle. Green edges represent other regular transfers.

\begin{center}
\includegraphics[scale = 0.5]{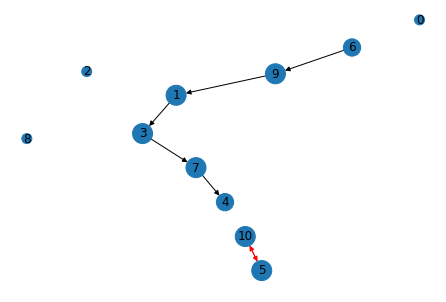}
\end{center}
\footnotesize \noindent Fig. 14.  Sale transactions for Azuki ID 5105;  Red edges are transactions in a wash sale cycle.

\medskip
\normalsize\par \textit{3) Repurchase Profit:}
\par A top repurchase gain identified by the algorithm is BAYC token ID 8099 (Fig. 15) showing that wallet \textbf{0xe4bc96b24e0bdf87b4b92ed39c1aef8839b090dd} sold the token for \$272K (60 ETH) to exit the cycle after previously repurchasing for \$174K (55 ETH) in August 2021. The large gain of nearly \$100K at the cycle’s exit indicates a high risk of the token being involved in price manipulation.

\begin{center}
\includegraphics[scale = 0.5]{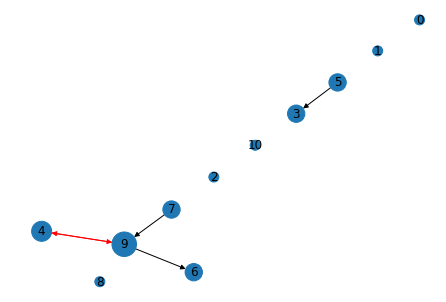}
\end{center}
\footnotesize \noindent Fig. 15.  Sale transactions for BAYC ID 8099. Red edges are transactions in a wash sale cycle

\medskip
\normalsize
\par \noindent \textit{B. Wallet Profit Analysis}
\par Within each collection, the sales profit was calculated at the wallet level to determine if any of the highest earners/losers also had wash trade behavior. The wallet level sales profit subtracts all outflows (purchases) from inflows (sales). It differs from the token level calculation as it effectively treats transfers as no value instead of passing through the previous value.
\par Analyzing the sales profits made in each collection for every participating wallet reveals that wallet address \textbf{0xd3fc6fec4b219c2d74b366fee6b585df71611533}, which made the most sales profit (\$8.8M) in the Bored Ape Yacht Club collection is a detected wash trader (Table IX). It is important to note that many of the sales from this address were transferred to this wallet from related addresses for no value. The pattern of an incoming transfer followed by a sale drives abnormally high wallet level profit while also demonstrating a common trait noticed in wash trade cycles. 
\par Furthermore, 3 wallets in the Azuki collection that participated in wash trading activity also ranked in the top 40 for sales profit amongst the 15 thousand wallets trading with Azuki tokens. They each benefitted from profits of \$566K, \$489K, and \$423K. These numbers demonstrate that wash traders often make larger profits than regular traders and further highlights how wash trading in the NFT market disrupts the market’s underlying integrity and security. 

\medskip
\par\noindent \textit{C. Limitations}
\par Information about rarity of a token or promotional efforts by the community that may have increased the value of a token were not taken into account when conducting case studies on specific tokens. Gas fees involved in each sale were not included in any calculations, and could potentially add more information and change some profit values and rankings.
\par Wallets that are flagged for potential involvement in wash trading due to having acted as an intermediary, although highly suspicious, could also be another unsuspecting buyer who was unaware of their role in a wash sale cycle. Due to the pseudo-anonymous nature of blockchain technology, there currently exists no method to automate the verification of the identity behind wallets in a cycle. The assumption of collusion was made based on anecdotal evidence from manual review, but will include false positives.
\par False negatives may also occur, as the algorithm is first alarmed by a sale transaction breaking the Wash Sale Rule. Attempts to artificially increase a token’s trading volume that never explicitly break the rule are not flagged. It is possible for market manipulators to regain possession of a token by creating a new wallet address for the repurchase. They can then either transfer it back into the original wallet’s ownership, or keep it in the new wallet that also belongs to them. As they never strictly “buy” it back, consequently never officially fulfilling the Wash Sale Rule, the algorithm does not mark this as a potential instance of wash trading. 
\par Wash traders could also wait until after a 30-day period before attempting a repurchase. Another project examining the timeline of wash trades in the NFT market has found that around 74.3\% of token trading cycles occur within the regulation threshold of 30 days [2]. The other 25.7\% of trading cycles that occur in a more than 30 day period, though legally less relevant,  also contain wash trading concerns and can be further examined.
\par To address these possibilities, the algorithm can be changed to initially run on transfer history as well as sales history, and without a time constraint of 30 days. It would then be able to mark transfers and sales for wash trades, but may operate less efficiently and take a looser definition of what it considers as wash trading.
\medskip
\par\noindent \textit{D. Future Use Cases}
\par Results from this analysis can be used in future attempts to construct a risk engine model. All three profit calculation methods provide a valuable signal for predictive machine learning models. Variables such as amount of wash trading activity and profits made from wash trading for each wallet address or within a token are applicable factors that affect their risk and reliability, and these values can be easily determined using the developed detection algorithm. Other features of interest involved in the analysis of wash sale risk include: 
\begin{itemize}
\item Average time between transactions in cycle
\item Minimum time between transactions in cycle
\item Address level risk indicators of each wallet in cycle
\item Gas fees spent on wash sale transactions 
\end{itemize}

Future iterations of the algorithm should also be run on a greater sample of collections to generate additional results and be more representative of wash trading in the NFT market.

\section*{V. References}

\end{multicols}

\end{document}